\documentclass[
superscriptaddress,
twocolumn,
showpacs,
bibnotes,
amsmath,
amssymb,
aps,
prl,
floatfix,
]{revtex4-2}

\usepackage{graphicx}
\usepackage{dcolumn}
\usepackage{bm}
\usepackage{cleveref}
\usepackage{silence}
\usepackage{ulem}
\usepackage{color}
\usepackage{soul}

\begin{document}


\title{Extraordinary Bulk Insulating Behavior in the Strongly Correlated Materials FeSi and FeSb$_2$} 
\author{Yun Suk Eo}
    \email{eohyung@umd.edu}
     \thanks{These authors contributed equally.}
    \affiliation{Maryland Quantum Materials Center and Department of Physics, University of Maryland, College Park, Maryland 20742, USA}

\author{Keenan Avers}
    \email[]{kavers@umd.edu}
    \thanks{These authors contributed equally.}
    \affiliation{Maryland Quantum Materials Center and Department of Physics, University of Maryland, College Park, Maryland 20742, USA}

\author{Jarryd A. Horn}
    \affiliation{Maryland Quantum Materials Center and Department of Physics, University of Maryland, College Park, Maryland 20742, USA}
    
\author{Hyeok Yoon}
    \affiliation{Maryland Quantum Materials Center and Department of Physics, University of Maryland, College Park, Maryland 20742, USA}
\author{Shanta Saha}
    \affiliation{Maryland Quantum Materials Center and Department of Physics, University of Maryland, College Park, Maryland 20742, USA}
    
\author{Alonso Suarez}
    \affiliation{Maryland Quantum Materials Center and Department of Physics, University of Maryland, College Park, Maryland 20742, USA}
    
\author{Michael S. Fuhrer}
    \affiliation{School of Physics and Astronomy, Monash University, Victoria 3800, Australia}
    \affiliation{ARC Centre of Excellence in Future Low-Energy Electronics Technologies, Monash University, Victoria 3800 Australia}
   
\author{Johnpierre Paglione}
    \affiliation{Maryland Quantum Materials Center and Department of Physics, University of Maryland, College Park, Maryland, USA}
    \affiliation{Canadian Institute for Advanced Research, Toronto, Ontario M5G 1Z8, Canada}
    \email{paglione@umd.edu}
    
\date{\today}
    
\begin{abstract}

  4$f$ electron-based topological Kondo insulators have long been researched for their potential to conduct electric current via protected surface states, while simultaneously exhibiting unusually robust insulating behavior in their interiors. To this end, we have investigated the electrical transport of the 3$d$-based correlated insulators FeSi and FeSb$_2$, which have exhibited enough similarities to their $f$ electron cousins to warrant investigation. By using a double-sided Corbino disk transport geometry, we show unambiguous evidence of surface conductance in both of these Fe-based materials. In addition, by using a 4-terminal Corbino inverted resistance technique, we extract the bulk resistivity as a function of temperature. Similar to topological Kondo insulator SmB$_6$, the bulk resistivity of FeSi and FeSb$_2$ are confirmed to exponentially increase by up to 9 orders of magnitude from room temperature to the lowest accessible temperature. This demonstrates that these materials are excellent bulk insulators, providing an ideal platform for studying correlated 2D physics.

\end{abstract}

\maketitle

 
\textit{Introduction.} 
The canonical Kondo insulators SmB$_6$ \cite{li2020emergent} and YbB$_{12}$ \cite{xiang2018quantum} have recently regained widespread interest following the identification of non-trivial band topology, studies of topological surface states, and the possible observations of charge-neutral fermions \cite{tan2015unconventional, hartstein2018fermi}.
It is now well established that the low-temperature plateau in electrical resistivity measurements of these materials originates from a surface channel contribution, consistent with the topological band inversion that is predicted to occur when Kondo hybridization opens a bulk band gap at the Fermi level \cite{TKI_Theory1, takimoto2011smb6}. Other examples of materials with apparent surface conduction have come to light, with low-temperature plateaus in resistivity arising despite apparent insulating behavior on cooling from room temperature. Most systems exhibit a resistivity increase of only a few factors at most before exhibiting saturation \cite{CaB6_Stankiewicz, YbB12_Ce343_CeNSn_takabatake1998and, Ce343_katoh1998crystal, Haen_TmSe, Malik_CeRhSb}. For these materials, a resistance plateau originating from surface conduction is unlikely as the low resistivity values of the plateaus would imply an unusually high sheet conductivity (using reasonable geometric factors). In contrast, a handful of correlated insulators including FeSi \cite{FeSi_Schlesinger}, FeSb$_2$ \cite{ FeSb2_Bentien}, and Ce$_3$Bi$_4$Pt$_3$ \cite{YbB12_Ce343_CeNSn_takabatake1998and} exhibit much larger (3-4 orders of magnitude) increases in resistivity before the plateau \cite{pickem2021resistivity} similar to the cases of SmB$_6$ and YbB$_{12}$.

In particular, the low-temperature resistivity saturation observed in the
iron-based correlated insulators FeSi and FeSb$_2$ have been suggested to originate from topological surface conducting states, as evidenced by transport \cite{fang2018evidence} and ARPES \cite{xu2020metallic} experiments. In contrast to the weakly correlated topological surface states, exotic phenomena that might be related to strong correlation characteristics such as surface magnetism (Zak phase in FeSi\cite{ThinFeSiZak_Ohtsuka}) and very low surface Fermi velocity ($v_{F}$ of $10^{3} - 10^4$ m/s in FeSb$_2$\cite{xu2020metallic}) have been reported in these two materials. 
Both materials exhibit striking similarities to the topological Kondo insulator SmB$_6$, in that their ground states are non-magnetic \cite{Jaccarino_ParaFeSi, SJOh_XPS_FESi,Zaliznyak_Neutron_FeSb2} despite having magnetic elements, and that they both have narrow band gaps \cite{acconductivityFeSi,perucchi2006optical} which in the case of SmB$_6$ arises due to Kondo physics \cite{mott1974rare, martin1979theory}. However, in FeSi and FeSb$_2$, this would require that the 3$d$ electrons participate in the gap opening instead of 4$f$ electrons, and since the 3$d$ electrons are less localized in nature, understanding the origin of the band gap is more difficult not as well agreed upon as in SmB$_6$.  Both FeSi \cite{fisk1996kondo, CastorFu_FeSi, MandrusFeSi_Thermodynamics} and FeSb$_2$ \cite{Petrovic_KondoInsulator} have been studied under the Kondo insulator framework, which involves a gap opening due to hybridization between a localized moment and a dispersive conduction band. However, there are other band calculations that show the band gap is between the 3$d$ multiplets (\cite{Mattheiss_LDA_FeSi} and \cite{mazin2021prediction, Chikina_BAndFeSb2, Tomczak_FeSb2_2010}). 
Overall, it is not confirmed that these materials share a common origin of bulk insulating behavior (e.g., Kondo effect), nor that topology plays a role in originating the apparent surface state conduction, raising the question on the nature of bulk and surface conduction in these materials.

While numerous studies of transport have been performed on SmB$_6$, the true bulk-insulating behavior was confirmed using a novel inverted resistance measurement technique \cite{InvertedResistance}. This technique, which accesses the bulk conductance by circumventing the dominating surface conduction channel via measurement of the voltage exterior of a Corbino disk \cite{eo2019transport}, revealed another remarkable feature of the bulk insulating behavior in SmB$_6$:  a thermally activated, ten orders-of-magnitude increase of the bulk resistivity on cooling from room temperature.

This measured exponential increase is in striking contrast to the behavior observed in conventional narrow-gap insulators or semiconductors, where the exponential rise of resistivity is typically terminated by extrinsic carriers from point defects or other disorders. For this reason, forming a truly insulating bulk in a semiconductor generally requires exceptionally pure materials, and indeed the ability to precisely control impurities is the foundation of the modern semiconductor industry. This is because conventional semiconductors obey the Mott criterion ($a_{B}N^{1/3} \approx 0.26$ \cite{mott1980metal}), where they transition into metal when the dopant concentration is higher than $10^{16} - 10^{17}$ cm$^{-3}$ (i.e. $\sim$0.0001 - 0.001\%).

In contrast, the insulating state in SmB$_6$ is robust to many orders of magnitude higher impurity density before the material transitions to a metal. This increase in bulk resistivity can still be seen in SmB$_6$ samples with up to several percent chemical substitution levels \cite{fuhrman2018screened, phelan2016chemistry, Phelan2014}. This is surprising considering only metals or s-wave superconductors typically allow such high levels of the substitution before transitioning to a distinct ground state. Since the insulating bulk of SmB$_6$ is so robust to (zero-dimensional) point defects, there is growing evidence that higher-order (one- or two-dimensional) defects such as dislocations are the leading type of disorder important for bulk conduction, and those defects are unconventional due to their topologically non-trivial nature \cite{FZFluxSmB6Inverted,xu2021intrinsic}. Given this unusual disorder and impurity response in bulk SmB$_6$, it is also of interest to investigate other correlated insulators for similar properties. Moreover, the characterization of such robust bulk-insulating systems provides an important foundation for the continuing surface states transport studies of FeSi and FeSb$_2$, and may even be the key technological advantage over more weakly correlated insulators.

In this study, we investigate the nature of bulk conductivity in the correlated insulators FeSi and FeSb$_2$, utilizing the inverted resistance technique to extract and compare their bulk resistivities. Confirmation of thermally activated bulk behavior in the low-temperature plateau region suggests these systems are truly bulk-insulating correlated materials, and that resistance saturation is due to surface conduction. The absence of bulk impurity conduction in both materials reveals another set of examples of extraordinary bulk insulation in a correlated insulator.

\begin{figure}
    \centering
    \includegraphics[scale=1.2]{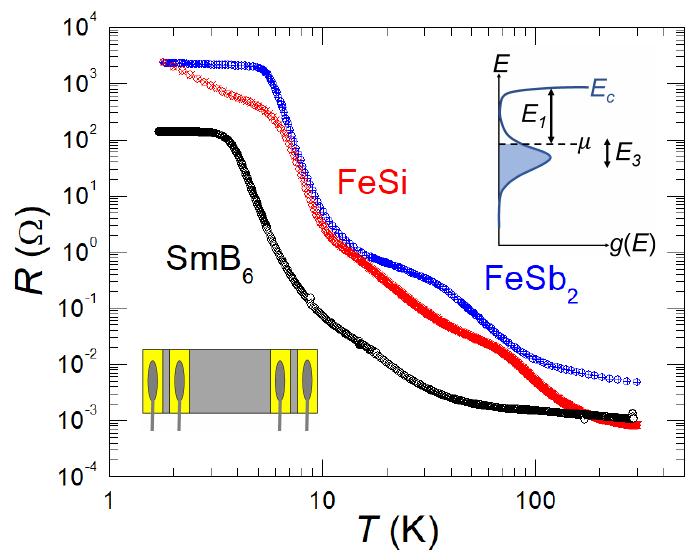}
    \caption{Typical resistance vs. temperature of FeSi (red) and FeSb$_2$ (blue) in comparison with SmB$_6$ (black). lower left inset: Resistances were measured using a conventional four-probe geometry.  Upper right inset: Schematic of an impurity band close to the conduction band. $E_{1}$ is the thermal activation energy from the chemical potential ($\mu$) to the nearest band (conduction band in the figure) and $E_{3}$ is the extrinsic thermal activation energy originating from the hopping conduction between impurity sites.}
    \label{fig:raw_resistance}
\end{figure}

\textit{Results}
We have prepared large single crystals ($\sim$ 5 mm polyhedra for FeSi and $\sim$ 2-3 mm polyhedra for FeSb$_2$), which easily allow for standard four-probe measurements, as depicted in the lower left inset of Fig.~(\ref{fig:raw_resistance}). The resistance ($R)$ vs. temperature ($T$) of FeSi and FeSb$_2$ in comparison with SmB$_6$ is shown in Fig.~(\ref{fig:raw_resistance}). The standard resistance of all three materials increases 5-6 orders of magnitude upon lowering the temperature, consistent with the previous literature \cite{SbFluxFeSi_Degiorgi, FeSiTransAnom_Glushkov, FeSiAndersonLocalized_Lisunov}. Most notably, all three standard resistances saturate at low temperatures. In SmB$_6$, this saturation below 4 K is due to a surface conduction layer, likely a gapless dispersion emerging from the non-trivial band topology. Recently, the existence of surface states has also been reported on FeSi and FeSb$_2$, using thickness-dependent transport \cite{fang2018evidence} and angle-resolved photoemission spectroscopy \cite{xu2020metallic}, respectively. These studies invite us to study the surface and hidden bulk conductivity of FeSi and FeSb$_2$. Indeed, later we verify the low-temperature saturation features in FeSi and FeSb$_2$ are of a surface origin, together with an estimation of the sheet resistance. 

\begin{figure}
    \centering
    \includegraphics[scale=1.1]{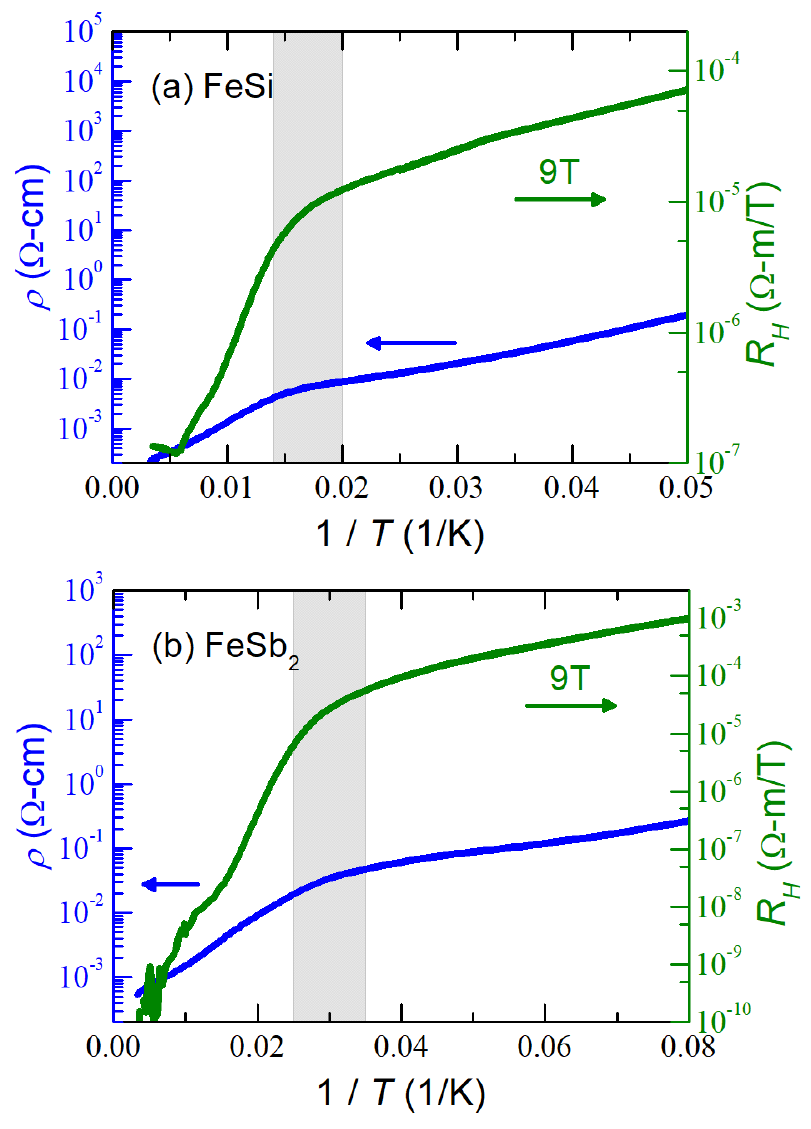}
    \caption{(color) Comparison of resistivity (blue and left axis) and 9 T Hall coefficient (green and right axis) at high temperatures focusing near the hump feature. The hump feature is shaded in gray. }
    \label{fig:Hall}
\end{figure}

But first, we comment on a hump feature at higher temperatures (SmB$_6$ at 15 K, FeSi at 50 K, FeSb$_2$ at 30 K). In SmB$_6$ this feature is weak, but it is much more pronounced in FeSi and FeSb$_2$, and can even be thought of as another saturation feature before the low-temperature surface one. To understand these hump features of FeSi and FeSb$_2$ more clearly, we also study the Hall coefficient as a function of temperature as well. As shown in Fig.~(\ref{fig:Hall}), we plot both resistivity and Hall coefficient as a function of inverse temperature. In this figure, we see that this feature exists in Hall effect and therefore it is likely due to a carrier density change by the shift of chemical potential (higher activation energy to lower activation energy). This change in activation energy has also been seen in the Hall effect of SmB$_6$, being consistent with the activation energy change from the middle of the gap to closer to the conduction band edge \cite{sluchanko1999nature}. The reason for this chemical potential shifting upon lowering the temperature is likely originating from a crossover from the intrinsic to the extrinsic regime (freeze out or ionization regime) \cite{Gorshunov} or band bending due to the surface states \cite{Rakoski}. Other temperature-dependent effects such as the bulk gap opening reported from ARPES\cite{Arita2008PRBFeSiARPES, xu2020metallic} and STM\cite{yang2021atomistic} studies may also play a role in the change in activation energy. We summarize the activation energy values in Table~(\ref{tab:activation_table}). The difference in the slope between resistivity and Hall indicates that mobility may also be a strong function of temperature, which requires in-depth follow-up studies. Lastly, it is important to note that the Hall coefficient of our FeSi and FeSb$_2$ lacks the Hall sign change, although sign changes of $R_H$ have been observed in FeSi previously \cite{Sun2014PRBFeSiThermoHallMagRes}. The Hall sign change is a feature commonly seen in high-quality $f$-electron systems attributed to skew scattering \cite{fert1987theory, coleman1985theory, rakoski2020investigation}. The absence of this sign change may reflect the high density of the extrinsic scattering centers \cite{rakoski2020investigation}. 

\begin{figure*}[t]
    \centering
     \includegraphics[scale=0.8]{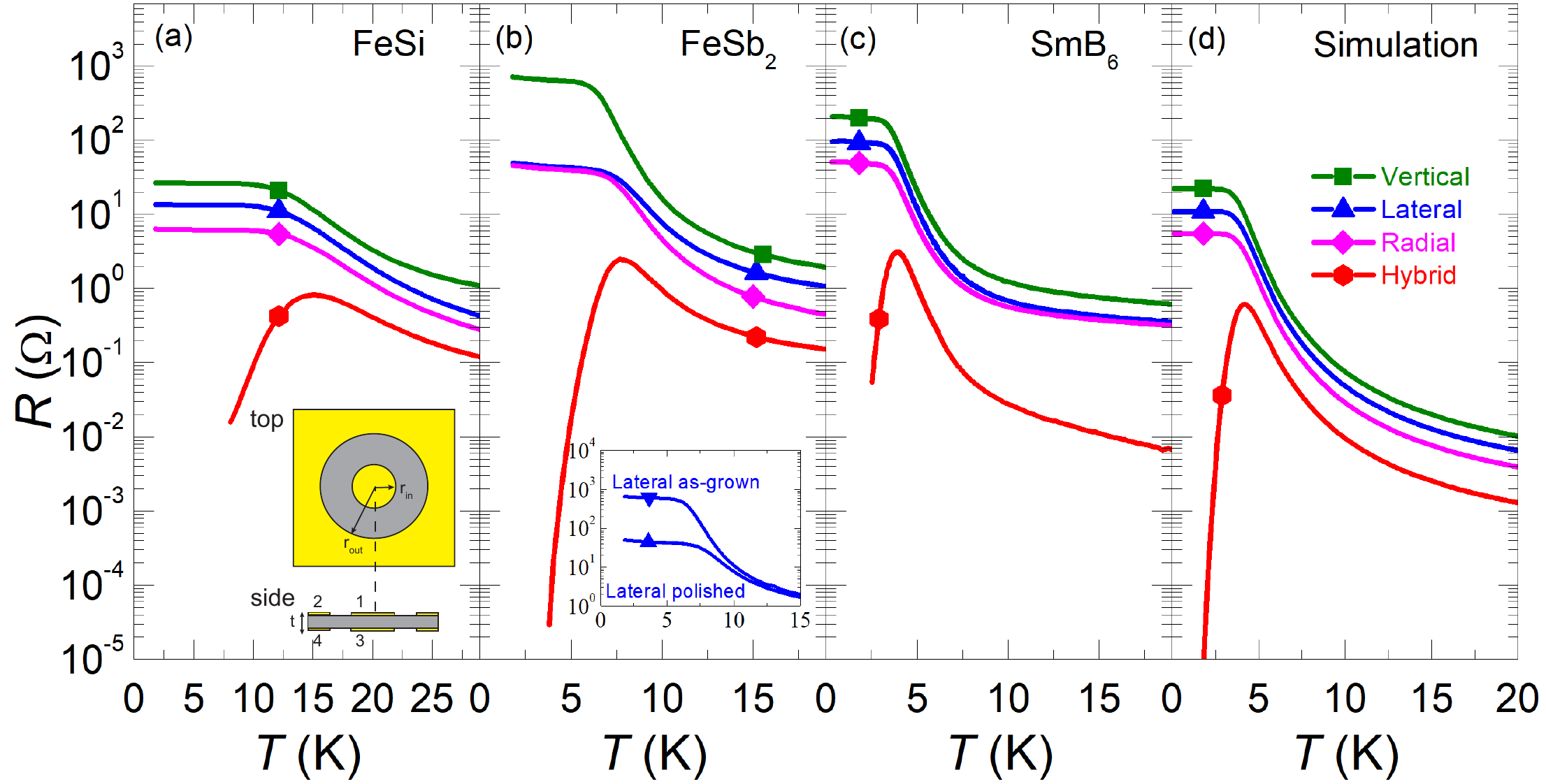}
    \caption{(Color) Surface conduction channel verification at low temperatures via employing coaxially-aligned Corbino disks (a) Inset: Schematic diagram of the Corbino disks on two opposite surfaces. Lateral (in blue): $R_{1,2}$ or $R_{3,4}$, Vertical (in green): $R_{1,3}$ while shorting 2 and 4, Radial (in magenta): $R_{1,2}$ while shorting 1 and 3; and 2 and 4. Hybrid (in red): $v_{1,2}/i_{3,4}$.  (a) $R$ vs $T$ measurement of FeSi (b) $R$ vs $T$ measurement of FeSb$_2$. Inset: The blue data with the upper triangle symbol is the lateral configuration of a Corbino disk on the polished surface, and the data with the lower triangle symbol is a lateral configuration measurement from an unpolished surface. (c) $R$ vs $T$ of SmB$_6$ from Ref. \cite{InvertedResistance} (d) Numerical (finite element analysis) demonstration of a cross over from insulating bulk to surface conduction upon lowering the temperature using a bulk activation energy is 3.5 meV and the sheet resistance is 100 $\Omega$. Details of the sample transport gemetry can be found in Appendix B.}    
    \label{fig:RawResistanceData}
\end{figure*}

We now show that the saturation of resistance at lower temperatures originates from the surface and not from the bulk. It is difficult to determine from a standard four-probe measurement if this saturation is of surface origin. Instead, we use a method introduced in Ref.~(\cite{InvertedResistance}), employing two Corbino disks coaxially aligned on two opposite surfaces. This allows us to measure what we call the lateral, hybrid, radial, and vertical resistances as described in the caption of Fig. \ref{fig:RawResistanceData}. If the resistance saturation is originating from the surface conduction and the bulk conduction is negligible, the lateral resistance is identical to a Corbino disk measurement on a two-dimensional electron gas or a thin film. Also, the radial and vertical resistance measurements are identical to two Corbino disk resistance measurements connected in parallel and in series, respectively. Most importantly, the hybrid measurement is an inverted resistance measurement, which essentially measures the voltage of the bulk current leaking out from a 2D Faraday cage (i.e., exterior of a Corbino disk). The hybrid resistance ($R_{\textrm{hybrid}}$) is given by 

\begin{equation}
    R_{\textrm{hybrid}} = C_{1}\frac{\sigma_{b}t}{\sigma_{s}^2}
    \label{eq:InvertedREsistance}
\end{equation}
where $C_1$ is a dimensionless geometric factor, $\sigma_{b}$ is the bulk conductance, $\sigma_{s}$ is the surface conductance, and $t$ is the thickness of the sample.
 This inverted (or hybrid) measurement is the key measurement of this study since it contains bulk conductivity information while the rest of the other measurements show saturation due to the weak temperature dependence of the surface resistance.

\begin{figure*}
    \centering
    \includegraphics[scale=1.2]{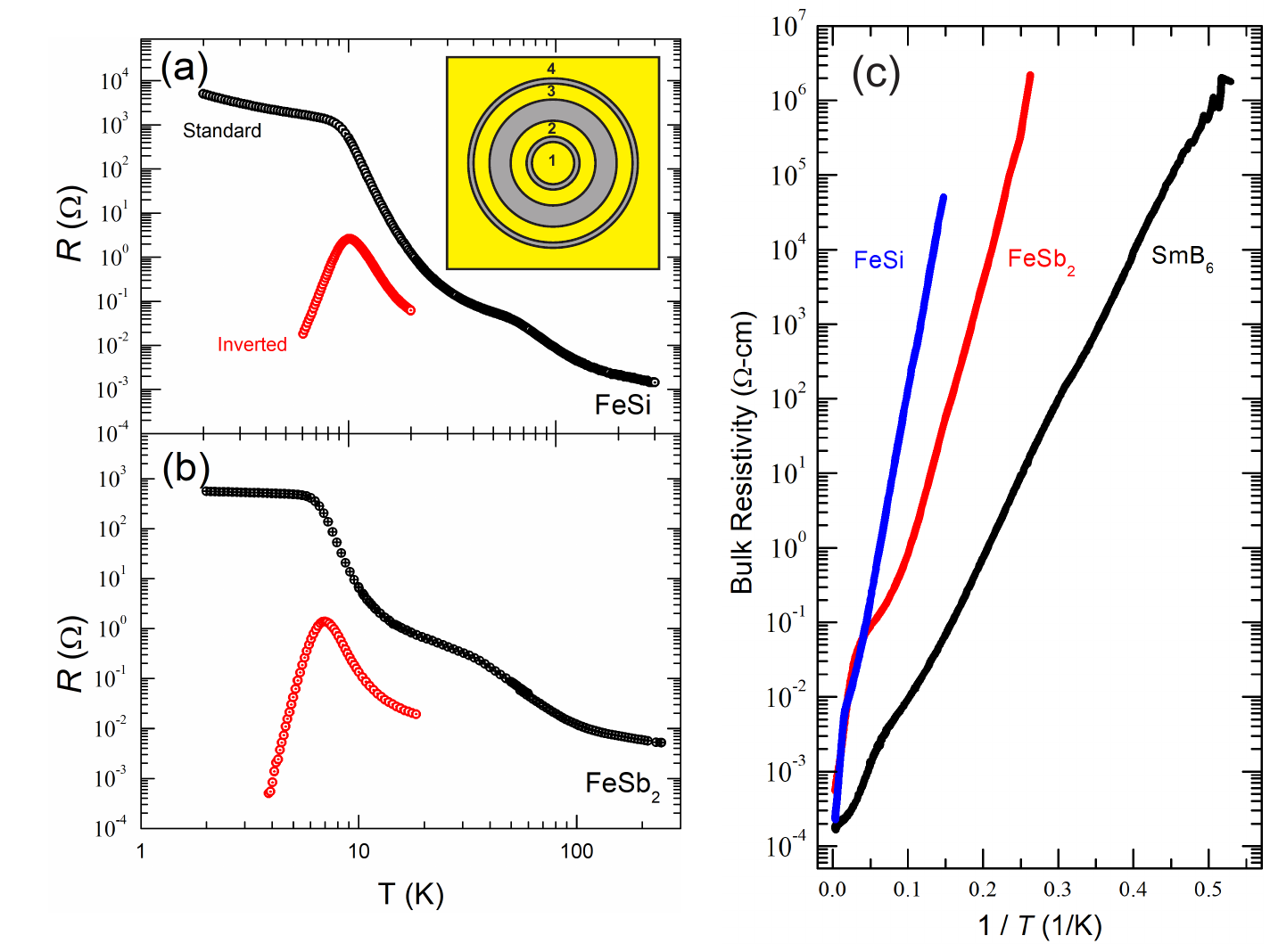}
    \caption{ (Color) Resistance measurement of a 4-terminal Corbino disk and bulk resistivity after the extraction process. (a) $R$ vs. $T$ of FeSi. Inset: Schematic diagram of the 4-terminal Corbino disk geometry. The conversion from resistance $R$ to sheet resistance $\rho^{2D}$ is $\rho^{2D} = 2\pi/\ln(3/2)\times R$. Standard resistance measurement configuration: $v_{1,4}/i_{2,3}$, Inverted resistance measurement configuration: $v_{1,2}/i_{4,3}$. (b) $R$ vs. $T$ of FeSb$_2$. (c) Bulk resistivity extraction result of FeSi (in blue) and FeSb$2$ (in red) in comparison with the previous SmB$_6$ (in black) report (from Ref.~\cite{eo2019transport}).  Details of the transport geometry and the bulk resistivity extraction process can be found in Appendix B and D, respectively. }
    \label{fig:bulkresistivity}
\end{figure*}

Experimental $R$ vs. $T$ from Corbino measurements are shown in Fig. \ref{fig:RawResistanceData} for FeSi (a) and FeSb$_2$ (b) in comparison with the previously reported SmB$_6$ (c) (from Ref. \cite{InvertedResistance}) and a numerical demonstration simulating a conducting surface and an insulating bulk (d). This experiment confirms that the resistance saturation in both FeSi and FeSb$_2$ originates from the surface conduction channel. First, the low temperature downturn of hybrid resistances consistent with Eq. \ref{eq:InvertedREsistance} (ie $\sigma_s\gg\sigma_b$t). In FeSi, both top and bottom surfaces were polished as identically as possible before patterning Corbino disks. In FeSb$_2$, in contrast, we polished only one surface and left the other surface in an as-grown condition before patterning the Corbino disks. The difference of the lateral resistance values between two opposite surfaces, as shown in the inset of Fig. \ref{fig:RawResistanceData} (b), is quite significant. The as-grown surface $R_{as grown}$ has an order of magnitude higher resistance value compared to the Corbino disk patterned on a roughly polished surface $R_{polished}$. This is similar to the previous SmB$_6$ report, showing evidence of subsurface crack conduction on a poorly prepared surface \cite{eo2020comprehensive, crivillero2021systematic}. Nevertheless, the radial and vertical measurements still show consistent behavior of the two channels connected in parallel and in series ($R_{vertical} = R_{polished} + R_{as grown} \approx R_{as grown}$ and $R_{radial}^{-1} = R_{as grown}^{-1} + R_{polished}^{-1} \approx R_{polished}$), again consistent with the surface state picture. For both FeSi and FeSb$_2$, we find that obtaining the same sheet resistance as the as-grown surface from finer polishing is much more challenging than in SmB$_6$, perhaps because the samples are much softer. Once the surface has been polished, the sheet resistance drops to a value that is almost an order of magnitude smaller. This can be either surface quality improvement or the creation of subsurface crack conduction channels, which requires further studies for clarification. Although in principle, we can extract the bulk resistivity with this setup, we prefer avoiding this change of sheet resistivity (or effectively changing it) since the inverted resistance will become a smaller value according to Eq.~(\ref{eq:InvertedREsistance}) (i.e, smaller $\rho_{s}$ becomes smaller $R_{inv}$ measurement). To this end, to measure the inverted resistance measurement for bulk resistivity extraction studies, we employ a four-terminal Cornino disk resistance measurement (as shown in the inset of Fig.~(\ref{fig:bulkresistivity})~(a)) patterned on an as-grown (unpolished) surface.

The bulk resistivity can be extracted by combining the information of the inverted resistance measurement and a standard four-terminal resistance measurement as shown in Fig.~(\ref{fig:bulkresistivity}) (a) and (b) (details are provided in the SI) \cite{InvertedResistance,eo2019transport}. The result of the bulk resistivity of FeSi and FeSb$_2$, compared to the SmB$_6$ \cite{eo2019transport}, is shown in Fig.~(\ref{fig:bulkresistivity})~(c). We note that the resistivity of FeSb$_2$ is an effective resistivity, where the current does not flow uniformly in an orthorhombic material. However, prior studies indicate the activation energies do not significantly differ depending on the directions \cite{bentien2007colossal}. We find that both FeSb$_2$ and FeSi show simple thermally activated behavior with nearly 8-9 orders of magnitude increase.

Lastly, we discuss the conducting surface channel. The resistance of FeSi from the standard Corbino measurement reads $R = 5.08$ k$\Omega$ in the surface-dominated regime. This corresponds to a sheet resistance of 78.7k$\Omega$, far exceeding $h/e^2$ which is the Mott-Ioffe-Regel (MIR) limit ($k_{F}l =1$) for a two-dimensional electron gas \cite{MIT_DasSarma}. This high value appears to rule out a metallic surface state emerging from a 3d strong topological insulator \cite{Nomura_scaling}, which should be protected against back-scattering and localization. For FeSb$_2$, the temperature dependence is much weaker and the sheet resistivity value does not exceed the MIR value. However, we note that $\sigma_{xx}$ and $\sigma_{yy}$ are not expected to be equal in general even on the 2d layer since the crystal is orthorhombic. Therefore, the sheet resistivity is an effective resistivity with  $\rho_{xx}$ and $\rho_{yy}$ not necessarily being an equal contribution. The details of the surface states in both FeSi and FeSb$_2$ will require future in-depth studies.

\textit{Discussion}

 Among correlated insulators, a robust insulating behavior of the bulk that increases exponentially by at least 8-9 orders of magnitude has only been seen in flux-grown SmB$_6$ \cite{eo2019transport}. In this study, we found two more materials behaving like this. Our finding is significant for future surface transport studies where interruption of the bulk channel is not acceptable. However, the detailed gap formation may be different in nature. The insulating gap of FeSi and FeSb$_2$ is likely to originate from the 3$d$ orbitals instead of the hybridization between a localized 4$f$ moment and a dispersive band. 
 
 In the historical literature of SmB$_6$ and FeSi, a saturation of resistance upon lowering the temperature has been interpreted as bulk metallic states by impurity conduction. In order to be valid, the authors have considered the Mott criterion and checked if the resistivity magnitude is consistent with a reasonable impurity concentration. In FeSi, the critical impurity density for the Mott criterion was reported to be $10^{18}$ cm$^{-3}$, and it was consistent with a resistivity value after increasing $\sim$ 5 orders of magnitude.

 We now find this saturation of resistance is a surface origin and the bulk continues to increase. The lowest temperature data point is limited by the performance of the electronics we used. Using the well-established transport theory of charged impurities in conventional semiconductors, the absence of a thermal activation energy change originating from hopping conduction (assisted by phonons) up to very high resistivity values suggests that FeSi and FeSb$_2$ has an impurity density that is lower than $5 \times 10^{-4}$ $\%$ and $2 \times 10^{-3}$ $\%$, respectively. This low impurity density is likely lower than the impurity level of our starting materials of the crystals. Either the unintentional impurities do not act as charged impurities (donors or acceptors) or our conventional understanding of impurities does not apply in these correlated insulators.

  It is worth mentioning different viewpoints of the bulk of SmB$_6$ which might be relevant to our Fe-based insulator studies. One speculation is that the Kondo gap may have a resemblance to an s-wave BCS superconducting gap whose existence is robust in the presence of a large number of impurities \cite{anderson1959theory}. Related to this view, it is worth mentioning that, in order to explain the experiments that support the experimental evidence of charge neutral fermions,  O. Erten $et$ $al$. views SmB$_6$ as a failed superconductor, where the order parameter does not have the topological stability to condensate into Cooper pairs, but it is instead a super dielectric \cite{erten_failedsuperconductor}. The resistivity and Hall coefficient temperature behavior has been explained by Rakoski $et$ $al$. without the presence of in-gap impurity states, but instead, band bending by the surface states being responsible for the detailed transport behavior \cite{Rakoski}. Alternatively, Souza $et$ $al$. \cite{metallicisland} and Jiao $et$ $al$. \cite{jiao2018magnetic} suggested the impurities are sealed off by metallic states by the topological nature. Lastly, B. Skinner explains the behavior by in-gap impurity states \cite{skinner2019properties}. In Skinner's model, if the dispersion can be approximated as a Mexican hat dispersion instead of a parabolic band, the insulator-to-metal transition is reserved until about $10^4$ times the doping density of the Mott criterion. Whether FeSi and FeSb$_2$ can also be understood within in these theoretical models needs to be investigated in future works. 

In conclusion, we have additionally discovered two robustly insulating correlated insulators: FeSi and FeSb$_2$, in the presence of surface states. We believe these additional material findings that have surface conduction channels with excellent insulation in the bulk will allow heterostructures for 2D flat band engineering.  

\begin{table*}
\centering
\begin{ruledtabular}
\begin{tabular}{l c c c c}
Sample/Measurement & $E_{high}$ (meV) ($T > T^{*}$)& $T^{*}$ (hump temperature)& $E_{low}$ (meV)\\
\hline
FeSi/Corbino & 25.71$\pm$0.0970 &  66 K & 10.82$\pm$0.0468 \\
FeSi/Hall & 44.52$\pm$0.178 & 55 K & 7.22$\pm$0.135 \\
FeSb$_2$/Corbino & 15.21$\pm$0.0833 & 35 K & 7.44$\pm$0.0290 \\
FeSb$_2$/Hall & 54.09$\pm$0.532 & 31 K & 6.26$\pm$0.197 \\
\end{tabular}
\end{ruledtabular}
\caption{\label{tab:activation_table} Activation energy fitting results of bulk resistivity and Hall coefficient. $T^{*}$ is the temperature at which the slope changes and shows as a hump in resistivity. The activation energies, $\Delta = E_{high}$ and $\Delta = E_{low}$, are estimated by fitting the functional form $\rho = \rho_{0}\exp (\Delta/k_{B}T)$ to the data.}
\end{table*}

\begin{acknowledgments}
We thank Ji-Hoon Park for the wire bonder assistance. We thank Ke-Jun Xu, Brian Skinner, Andriy Nevidomskyy, Shouvik Sur, and Onur Erten for the discussions.

\end{acknowledgments} 

\bibliography{bibliography}

\end{document}